\input harvmac
\def\figflag{I}
\noblackbox

\font\cmss=cmss10 \font\cmsss=cmss10 at 7pt
 \def\inbar{\,\vrule height1.5ex width.4pt depth0pt}
\def\IZ{\relax\ifmmode\mathchoice
{\hbox{\cmss Z\kern-.4em Z}}{\hbox{\cmss Z\kern-.4em Z}} {\lower.9pt\hbox{\cmsss Z\kern-.4em Z}}
{\lower1.2pt\hbox{\cmsss Z\kern-.4em Z}}\else{\cmss Z\kern-.4em Z}\fi}
\def\IB{\relax{\rm I\kern-.18em B}}
\def\IC{{\relax\hbox{$\inbar\kern-.3em{\rm C}$}}}
\def\ID{\relax{\rm I\kern-.18em D}}
\def\IE{\relax{\rm I\kern-.18em E}}
\def\IF{\relax{\rm I\kern-.18em F}}
\def\IG{\relax\hbox{$\inbar\kern-.3em{\rm G}$}}
\def\IGa{\relax\hbox{${\rm I}\kern-.18em\Gamma$}}
\def\IH{\relax{\rm I\kern-.18em H}}
\def\II{\relax{\rm I\kern-.18em I}}
\def\IK{\relax{\rm I\kern-.18em K}}
\def\IR{\relax{\rm I\kern-.18em R}}

\def\tfig#1{Figure~\the\figno\xdef#1{Figure~\the\figno}\global\advance\figno by1}
\def\figI{I}
%
\newdimen\tempszb \newdimen\tempszc \newdimen\tempszd \newdimen\tempsze
\ifx\figflag\figI
\input epsf
%
%
%
%
%
\def\ifigure#1#2#3#4{
\midinsert \vbox to #4truein{\ifx\figflag\figI \vfil\centerline{\epsfysize=#4truein\epsfbox{#3}}\fi}
\baselineskip=12pt \narrower\narrower\noindent{\bf #1:} #2
\endinsert
}

\lref\WittenNF{
  E.~Witten,
    ``Dynamical Breaking Of Supersymmetry,''
      Nucl.\ Phys.\  B {\bf 188}, 513 (1981).
        }

\lref\AffleckXZ{
  I.~Affleck, M.~Dine and N.~Seiberg,
    ``Dynamical Supersymmetry Breaking In Four-Dimensions And Its
      Phenomenological Implications,''
        Nucl.\ Phys.\  B {\bf 256}, 557 (1985).
          }

\lref\IntriligatorCP{
  K.~Intriligator and N.~Seiberg,
    ``Lectures on Supersymmetry Breaking,''
      arXiv:hep-ph/0702069.
        }

\lref\VerlindeJR{
  H.~Verlinde and M.~Wijnholt,
    ``Building the standard model on a D3-brane,''
      JHEP {\bf 0701}, 106 (2007)
        [arXiv:hep-th/0508089].
      }

\lref\PoppitzVD{
  E.~Poppitz and S.~P.~Trivedi,
    ``Dynamical supersymmetry breaking,''
      Ann.\ Rev.\ Nucl.\ Part.\ Sci.\  {\bf 48}, 307 (1998)
        [arXiv:hep-th/9803107].
          }

\lref\VerlindeBC{
  H.~Verlinde,
    ``On metastable branes and a new type of magnetic monopole,''
      arXiv:hep-th/0611069.
    }

\lref\AmaritiAM{
  A.~Amariti, L.~Girardello and A.~Mariotti,
    ``Meta-stable $A_n$ quiver gauge theories,''
      arXiv:0706.3151 [hep-th].
	}

\lref\ShadmiJY{
  Y.~Shadmi and Y.~Shirman,
    ``Dynamical supersymmetry breaking,''
      Rev.\ Mod.\ Phys.\  {\bf 72}, 25 (2000)
        [arXiv:hep-th/9907225].
          }

\lref\WittenBN{
  E.~Witten,
    ``Non-Perturbative Superpotentials In String Theory,''
      Nucl.\ Phys.\  B {\bf 474}, 343 (1996)
        [arXiv:hep-th/9604030].
      }

\lref\BeckerKB{
  K.~Becker, M.~Becker and A.~Strominger,
    ``Five-Branes, Membranes And Nonperturbative String Theory,''
      Nucl.\ Phys.\  B {\bf 456}, 130 (1995)
        [arXiv:hep-th/9507158].
      }

\lref\DineGM{
  M.~Dine, J.~L.~Feng and E.~Silverstein,
    ``Retrofitting O'Raifeartaigh models with dynamical scales,''
      Phys.\ Rev.\  D {\bf 74}, 095012 (2006)
        [arXiv:hep-th/0608159].
          }

\lref\IntriligatorDD{
  K.~Intriligator, N.~Seiberg and D.~Shih,
    ``Dynamical SUSY breaking in meta-stable vacua,''
      JHEP {\bf 0604}, 021 (2006)
        [arXiv:hep-th/0602239].
      }

\lref\BlumenhagenMU{
  R.~Blumenhagen, M.~Cvetic, P.~Langacker and G.~Shiu,
    ``Toward realistic intersecting D-brane models,''
      Ann.\ Rev.\ Nucl.\ Part.\ Sci.\  {\bf 55}, 71 (2005)
        [arXiv:hep-th/0502005].
      }

\lref\FloreaSI{
  B.~Florea, S.~Kachru, J.~McGreevy and N.~Saulina,
    ``Stringy instantons and quiver gauge theories,''
      JHEP {\bf 0705}, 024 (2007)
        [arXiv:hep-th/0610003].
      }
\lref\BlumenhagenXT{
  R.~Blumenhagen, M.~Cvetic and T.~Weigand,
    ``Spacetime instanton corrections in 4D string vacua - the seesaw mechanism
      for D-brane models,''
        Nucl.\ Phys.\  B {\bf 771}, 113 (2007)
      [arXiv:hep-th/0609191].
        }
\lref\IbanezDA{
  L.~E.~Ibanez and A.~M.~Uranga,
    ``Neutrino Majorana masses from string theory instanton effects,''
      JHEP {\bf 0703}, 052 (2007)
        [arXiv:hep-th/0609213].}
\lref\ArgurioQK{
  R.~Argurio, M.~Bertolini, S.~Franco and S.~Kachru,
    ``Metastable vacua and D-branes at the conifold,''
      JHEP {\bf 0706}, 017 (2007)
        [arXiv:hep-th/0703236].
      }
\lref\GanorPE{
  O.~J.~Ganor,
    ``A note on zeroes of superpotentials in F-theory,''
      Nucl.\ Phys.\  B {\bf 499}, 55 (1997)
        [arXiv:hep-th/9612077].
      }

\lref\UrangaVF{
  A.~M.~Uranga,
    ``Brane configurations for branes at conifolds,''
      JHEP {\bf 9901}, 022 (1999)
        [arXiv:hep-th/9811004].
          }

\lref\GiudiceBP{
  G.~F.~Giudice and R.~Rattazzi,
    ``Theories with gauge-mediated supersymmetry breaking,''
      Phys.\ Rept.\  {\bf 322}, 419 (1999)
        [arXiv:hep-ph/9801271].
      }

\lref\KawanoRU{
T.~Kawano, H.~Ooguri and Y.~Ookouchi,
``Gauge Mediation in String Theory,"
arXiv:0704.1085 [hep-th].
}

\lref\DiaconescuPC{
D.~E.~Diaconescu, B.~Florea, S.~Kachru and P.~Svrcek,
``Gauge-mediated supersymmetry breaking in string compactifications,"
JHEP {\bf 0602}, 020 (2006)
[arXiv:hep-th/0512170].
}

\lref\FrancoII{
  S.~Franco, A.~Hanany, D.~Krefl, J.~Park, A.~M.~Uranga and D.~Vegh,
    ``Dimers and Orientifolds,''
      arXiv:0707.0298 [hep-th].
        }

\lref\AharonyPR{
  O.~Aharony and S.~Kachru,
    ``Stringy Instantons and Cascading Quivers,''
      arXiv:0707.3126 [hep-th].
        }

\lref\HaackCY{
  M.~Haack, D.~Krefl, D.~Lust, A.~Van Proeyen and M.~Zagermann,
    ``Gaugino condensates and D-terms from D7-branes,''
      JHEP {\bf 0701}, 078 (2007)
        [arXiv:hep-th/0609211].
      }

\lref\IbeKM{
  M.~Ibe and R.~Kitano,
  ``Sweet Spot Supersymmetry,''
  arXiv:0705.3686 [hep-ph].
}

\lref\Feng{J. Feng et al, work in progress.}

\lref\ShihAV{
  D.~Shih,
  ``Spontaneous R-symmetry breaking in O'Raifeartaigh models,''
  arXiv:hep-th/0703196.
}

\lref\KachruGS{
  S.~Kachru, J.~Pearson and H.~L.~Verlinde,
    ``Brane/flux annihilation and
    the string dual of a non-supersymmetric  field
      theory,''
        JHEP {\bf 0206}, 021 (2002)
      [arXiv:hep-th/0112197].
        }

\lref\AganagicEX{
  M.~Aganagic, C.~Beem, J.~Seo and C.~Vafa,
    ``Geometrically induced metastability and holography,''
      arXiv:hep-th/0610249.
    }

\lref\MaloneyRR{
  A.~Maloney, E.~Silverstein and A.~Strominger,
    ``De Sitter space in noncritical string theory,''
      arXiv:hep-th/0205316.
    }

\lref\SaltmanJH{
  A.~Saltman and E.~Silverstein,
    ``A new handle on de Sitter compactifications,''
      JHEP {\bf 0601}, 139 (2006)
        [arXiv:hep-th/0411271].
      }

\lref\DouglasTU{
  M.~R.~Douglas, J.~Shelton and G.~Torroba,
    ``Warping and supersymmetry breaking,''
      arXiv:0704.4001 [hep-th].
    }

\lref\HeckmanWK{
  J.~J.~Heckman, J.~Seo and C.~Vafa,
    ``Phase Structure of a Brane/Anti-Brane System at Large N,''
      arXiv:hep-th/0702077.
    }

\lref\HeckmanUB{
  J.~J.~Heckman and C.~Vafa,
    ``Geometrically Induced Phase Transitions at Large N,''
      arXiv:0707.4011 [hep-th].
    }

\lref\KachruAW{
  S.~Kachru, R.~Kallosh, A.~Linde and S.~P.~Trivedi,
    ``De Sitter vacua in string theory,''
      Phys.\ Rev.\  D {\bf 68}, 046005 (2003)
        [arXiv:hep-th/0301240].
      }

\lref\KlebanovHB{
  I.~R.~Klebanov and M.~J.~Strassler,
  ``Supergravity and a confining gauge theory: Duality cascades and
  $\chi$SB-resolution of naked singularities,''
  JHEP {\bf 0008}, 052 (2000)
  [arXiv:hep-th/0007191].
}

\lref\GarciaEtxebarriaRW{
  I.~Garcia-Etxebarria, F.~Saad and A.~M.~Uranga,
  ``Local models of gauge mediated supersymmetry breaking in string theory,''
  JHEP {\bf 0608}, 069 (2006)
  [arXiv:hep-th/0605166].
}


\Title {\vbox{\baselineskip12pt\hbox{SLAC-PUB-12698} \hbox{SU-ITP-07/11} \hbox{WIS/12/07-AUG-DPP}}}
{\vbox{
\centerline{Simple Stringy Dynamical SUSY Breaking}} } \centerline{Ofer Aharony$^{1,2},$ Shamit Kachru$^{2}$ and
Eva Silverstein$^{2}$ }

\bigskip
\centerline{$^{1}$Department of Particle Physics} \centerline{Weizmann Institute of Science} \centerline{Rehovot
76100, Israel}
\smallskip
\medskip
\centerline{$^{2}$ Department of Physics and SLAC} \centerline{Stanford University} \centerline{Stanford, CA
94305, USA}
\medskip
\bigskip
\medskip
\noindent We present simple string models which dynamically break
supersymmetry without non-Abelian gauge
dynamics.  The Fayet model, the Polonyi model, and the O'Raifeartaigh
model each arise from D-branes at a specific type of
singularity. D-brane instanton effects generate the requisite
exponentially small scale of supersymmetry breaking.

\Date{August 2007}


\newsec{Introduction}

Dynamical supersymmetry breaking (DSB) is a promising candidate
solution to the hierarchy problem \WittenNF. Many field theories which
dynamically break supersymmetry have been discovered (see
\refs{\AffleckXZ,\PoppitzVD,\ShadmiJY,
\IntriligatorCP} for reviews).  In each of these examples,
non-Abelian gauge dynamics plays a crucial role. In general, the
constructions are rather complicated, though they have become simpler
over the years \IntriligatorCP.

One way to dynamically break supersymmetry (SUSY) in string theory is
to embed a non-Abelian gauge theory which dynamically breaks
supersymmetry into the low-energy spectrum.  Here, we propose an
alternative.  We find simple D-brane theories which dynamically break
supersymmetry after including D-brane instanton effects.
The low-energy theories are Fayet, Polonyi or
O'Raifeartaigh models.  The terms in the superpotential which are
responsible for supersymmetry breaking arise
due to stringy D-instanton generated perturbations,
which have recently been investigated in
\refs{\BlumenhagenXT,\HaackCY,\IbanezDA,\FloreaSI} and many
subsequent papers.\foot{Early work
on similar instanton effects appears in \refs{\BeckerKB,\WittenBN,\GanorPE}.}
Non-Abelian gauge dynamics plays no role, and the SUSY breaking
``hidden sectors'' are extremely modest in size, including a single
Abelian gauge field with two charged chiral multiplets or even more
minimal field content. One can view our results as indicating that
stringy instantons make retrofitting of simple supersymmetry-breaking
models \DineGM\ a natural feature of D-brane constructions.  Because
of the importance of the stringy instanton effect, in these models the
brane construction plays a more fundamental role than just serving as
a way to embed a known low-energy field theory mechanism into string
theory.\foot{Some other papers which study
stringy mechanisms to break supersymmetry
using
systems of branes, anti-branes and fluxes are
\refs{\KachruGS\MaloneyRR\KachruAW\SaltmanJH\AganagicEX\VerlindeBC
\HeckmanWK\DouglasTU-\HeckmanUB}.}

In \S2, we describe the simplest models we have found. All of these
models can arise from D-branes at a specific singularity, which can be
chosen to be an
orientifold of an orbifold of the conifold. In \S3, we
briefly discuss the prospects for making fully realistic models using our
SUSY breaking hidden sectors. The construction of complete models
utilizing our SUSY breaking models as hidden sectors is left for
future work.

\newsec{Some Simple Models}

In this section we present
simple D-brane theories where stringy instanton effects yield vacua with
exponentially small SUSY breaking scale.  The low-energy theories are
Fayet, Polonyi and O'Raifeartaigh models.

\subsec{The Fayet Model}

The Fayet model consists of a $U(1)$ gauge field coupled with strength
$e$ to charged chiral multiplets $\Phi_{\pm}$ with equal and opposite
charges and canonical K\"ahler potential.  The superpotential is
\eqn\fmodel{W = m \Phi_+ \Phi_- ~,}
so the F-term equations for supersymmetric vacua require the scalar
components to satisfy $\phi_{\pm} = 0$. The D-term constraints require
supersymmetric vacua to satisfy
\eqn\dterm{|\phi_+|^2 - |\phi_-|^2 = r,}
where $r$ is the Fayet-Iliopoulos (FI) D-term for the Abelian gauge
symmetry.

For generic values of the FI term $r \neq 0$, the F-term equations and
\dterm\ cannot be simultaneously satisfied.  The energy grows without
bound at infinity in field space, so this model has a stable ground
state which spontaneously breaks supersymmetry.  Specifically, for $r
\gg m^2/(2e^2)$, the minimum of the scalar potential is at
$|\phi_+|^2=r-m^2/(2e^2)\simeq r$, $\phi_-=0$, and the breaking of supersymmetry is
dominated by the $F$-term
\eqn\Fscale{F_{\Phi_-}\simeq m \sqrt{r}.}
%
We will now exhibit a simple brane
realization of this model, with an exponentially small supersymmetry
breaking scale obtained by generating $m$ from a stringy instanton
effect.

The basic idea is as follows. We can realize the theory described
above as a quiver gauge theory, arising at low energies on D-branes
probing a non-compact singular Calabi-Yau space in type IIB string
theory (or, equivalently, from D-branes stretched between NS-branes
in type IIA string theory). The relevant quiver for us is quite simple
and could potentially arise from many singularities; it appears below
in \tfig\figone. It has two $U(r)$ nodes of rank $r=1$ and one
$USp(r)$ node of rank
$r=0$. In the geometrical language,
the space locally contains two 2-cycles on which space-filling
5-branes (often called ``fractional branes") are wrapped, and another
2-cycle ${\cal C}$ which is not wrapped by a 5-brane.  There are two
chiral multiplets arising from open strings between the 5-branes, with
charges $(\pm 1, \mp 1)$ under the $U(1) \times U(1)$ gauge group. The
superpotential is zero perturbatively.  A Euclidean D1-brane wrapped
on ${\cal C}$ contributes an instanton effect with precisely the right
zero-mode structure to generate the superpotential \fmodel; this
cannot be interpreted as an ordinary field-theoretic instanton, since
there is no field theory associated with this cycle, and no
non-Abelian gauge dynamics is required for the effect. $m$ and $r$ are
fixed parameters at the level of the non-compact system since they
arise from non-normalizable modes.\foot{In a compact model with finite
four dimensional Planck scale, these modes become dynamical. Then, as
with all proposals for dynamical supersymmetry breaking in string
theory, one must stabilize the closed string moduli which control the
scales of the gauge theory.}

\ifigure\figone{The quiver diagram that leads to the Fayet model. The
first, square, node corresponds to a $USp(r_1)$ group, while the
circular nodes correspond to $U(r_i)$ groups. For our application we
need to have $r_2=r_3=1$, and $r_1=0$ (this is the node wrapped by the
D-instanton); the bifundamentals connecting node 1 and node 2 are then
Ganor strings.}
{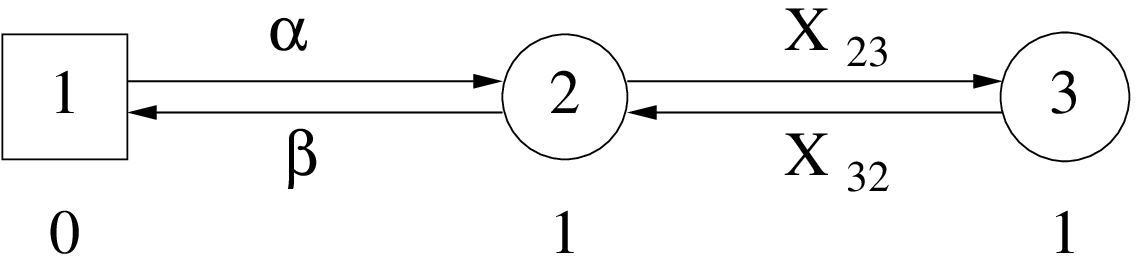}{1.0}

Concretely, we can obtain the simple subquiver in \figone, as well as
a generalization relevant for gauge mediation to be discussed in \S3,
starting from the singular geometries
\eqn\singis{(xy)^n = zw~.}
These are $\IZ_n$
orbifolds of the conifold, studied in \UrangaVF.\foot{The quivers we use
can probably be obtained from many other singularities as well.}
The quivers
describing the effective gauge theories living on D3 and D5-branes at
these singularities have $2n$ $U(r_i)$ nodes with bifundamentals
$X_{i,i+1},X_{i+1,i}$ going each way between adjacent nodes, as in the
left-hand side of
\tfig\figtwo, and with a superpotential
\eqn\quivsup{W = h \sum_{i=1}^{2n} ~(-1)^{i}~X_{i,i+1} X_{i+1,i+2}
X_{i+2,i+1} X_{i+1,i}~.}
Specific orientifolds of this theory which lead to interesting stringy instanton effects were described in
\refs{\ArgurioQK,\FrancoII}. In the case where the quiver nodes are occupied by space-filling wrapped branes,
these modify the field content such that nodes $1$ and $n+1$ correspond to symplectic gauge groups instead of
unitary groups, while the remaining $U(r_i)$ nodes are pairwise identified by the obvious reflection symmetry.
The identification of node 1 with itself by the orientifold is important because it reduces the number of
fermion zero modes on the Euclidean D1-brane wrapping the corresponding cycle ${\cal C}$ to the two that are
required for a contribution to the space-time superpotential. The T-dual type IIA string description of the
branes at this orientifolded orbifolded conifold is shown in \tfig\figthree.

\ifigure\figtwo{The quiver gauge theories of the orbifolded conifold and
of its orientifold for $n=3$. The circular nodes have $U(r_i)$
gauge groups, and the square nodes have $USp(r_i)$ groups.
More generally there are $2n$ nodes before orientifolding and $n+1$
nodes after orientifolding.}
{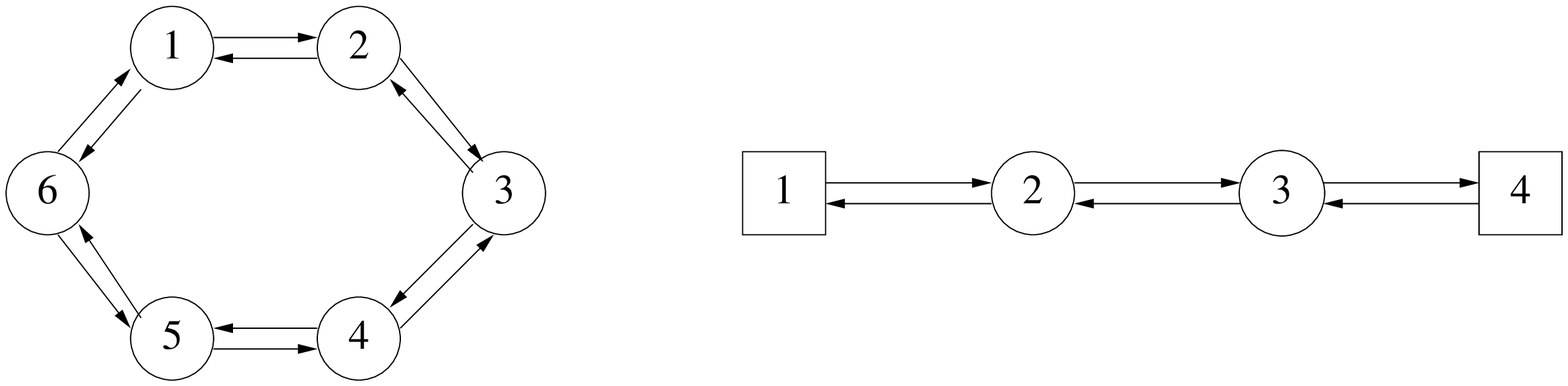}{1.5}

The model we are interested in arises when we have $n \geq 3$, and we have single (space-filling) branes on
nodes 2 and 3 ($r_2 = r_3 = 1$), and vanishing occupation numbers elsewhere.  The tree-level superpotential
\quivsup\ vanishes in this case. The D-instanton wrapping node 1 has bifundamental fermionic ``Ganor strings''
$\alpha$ and $\beta$ stretching to node 2 \refs{\GanorPE,\ArgurioQK} (see \figone).  These modes have a coupling
analogous to \quivsup\ to the fields $X_{23}, X_{32}$; performing the path integral over $\alpha$ and $\beta$
then generates a superpotential \refs{\ArgurioQK, \AharonyPR}
\eqn\instsup{W = \Lambda_{1} X_{23}X_{32}~,}
where $\Lambda_1$ is the instanton action controlled by the size of
node 1 in the geometry, and it can naturally be exponentially
small.\foot{For $n=3$ there would be a similar contribution arising
also from node 4.}

\ifigure\figthree{The T-dual type IIA brane configuration for our
Fayet model when it is embedded in
the $n=3$ orientifold. NS branes stretch in the 012345 directions,
NS' branes in the 012389 directions, and D4 branes stretch in the
01236 directions. The O6 planes extend along the 01237 directions,
and lie at a 45 degree angle with respect to the 45 and 89
planes. The $x^6$ direction is compact and becomes an interval after
the orientifolding.}
{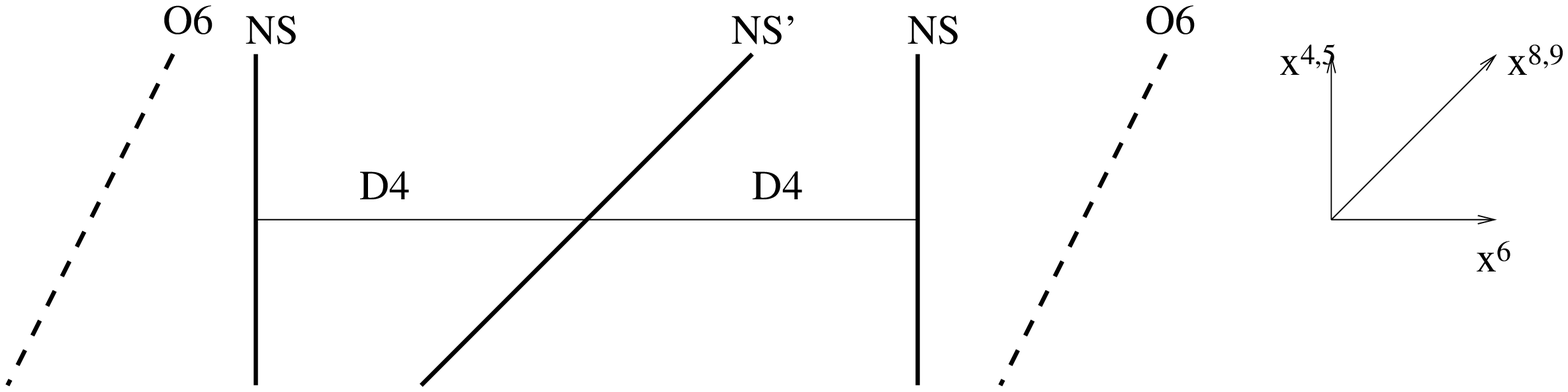}{1.8}

The sum of the $U(1)$'s associated to nodes 2 and 3 acts trivially on
all fields and decouples.  The low-energy theory consists of a single
$U(1)$ gauge field (the difference of the $U(1)$'s at the two nodes),
with $X_{23}$ and $X_{32}$ carrying equal and opposite charges. This
$U(1)$ does not decouple at low energies, because its renormalization
group running stops below the scale of the mass of the charged
fields. Thus, we obtain precisely the Fayet model, with
$\Phi_+=X_{23}$, $\Phi_-=X_{32}$, and with the parameter $m$ of \fmodel\
having been dynamically generated by a D-instanton. For generic
choices of the FI term $r$ (which is a non-normalizable mode in the
non-compact geometry), this model breaks supersymmetry at an
exponentially low scale $F\sim \Lambda_1 \sqrt{r}$
\Fscale.  This can be considered a retrofitted Fayet model, in the
spirit of \DineGM. However, no non-Abelian gauge dynamics is invoked
in the retrofitting; it is automatically implemented by string
theory. In the type IIA language of \figthree, the FI term corresponds
to the $x^7$ position of the NS' between the two
D4-branes.

The effective action of the model described above (and of the models
we will discuss below) will in general be
corrected by higher-dimension operators.  These generically shift the
location of the vacuum slightly, and can also introduce a
supersymmetric vacuum elsewhere in field space, rendering the
SUSY breaking vacuum metastable.

\subsec{The Polonyi model}

An even simpler model of SUSY breaking is the Polonyi model. This is
the theory of a single chiral superfield with superpotential
\eqn\polonyi{W = \mu^2 X~.} $F_X = \mu^2$ provides the order parameter
of SUSY breaking.  At tree level, this model has a flat direction.
The existence of a stable non-SUSY vacuum at $X=0$ depends on the sign of the
leading quartic correction to the K\"ahler potential
\eqn\polKah{K = X^\dagger
X + {c\over M_*^2} (X^\dagger X)^2 +
\cdots.}
$M_*$ denotes the scale of high-energy physics which has been integrated out and corrects $K$. For one
sign of $c$ there is a stable vacuum, and for the other the theory
runs away to large values of $X$.  In any given completion of the
Polonyi model by a larger field theory or string theory, there will be
some corresponding value of $c$.

In fact, a particularly simple completion manifesting a stable vacuum
is provided by the Fayet model discussed above, which reduces to the
Polonyi model in a limit.  At the level of the field theory model, as
$r$ grows large, with $m\sqrt{r}\equiv \mu^2$ fixed, the $U(1)$ under
which $\Phi_\pm$ are charged becomes very massive along with
$\Phi_+$. The remaining $U(1)$ is free and decouples as before. The
low-energy theory therefore reduces to a free $U(1)$ theory with a singlet
$X=\Phi_-$ that has mass squared $2m^2=2\mu^4/r$ (which goes to
zero in our limit) and a linear
superpotential $W=\mu^2 X$ as in \polonyi. In the string construction
realizing this model, we must keep $r$ smaller than the string scale
to avoid introducing new degrees of freedom; this still leaves a
regime where the low energy effective theory is the Polonyi model with
a locally stable minimum.

In the brane construction of \figthree, turning on a large FI term
corresponds to moving an NS' brane far away in the $x^7$
direction. One could also obtain the Polonyi model directly, with a
dynamically generated small scale $\mu^2$, by considering the brane
configuration without this NS' brane, such that we have a single
D4-brane stretched between two parallel NS5-branes (with orientifolds
as in \figthree). This corresponds to the quiver shown in
\tfig\figfour, with $r_2=1$ and $r_1$ vanishing.
In the brane language, the field $X$ arises as the translation mode
of the D4 along the $x^4$ and $x^5$ directions, and the stringy
instanton is in this language the Euclidean D0 brane wrapping the
interval between the NS 5-brane and the O6-plane. The Ganor strings
now have an action of the form $S = \alpha \beta X$, so that this
instanton gives precisely a superpotential of the form \polonyi.

Instead of moving away an NS' brane along $x^7$ as described above,
one can also obtain this brane configuration from the one in
\figthree\ by moving the NS' in the
$x^6$ direction so that it trades places with an NS brane
(annihilating the D4-branes ending on it in the process).  Such a
position-switching process actually happens during the renormalization
group cascade \KlebanovHB\ which arises for branes with large
occupation numbers at the singularities described in the previous
subsection.  As described in \AharonyPR, the cascade steps lead to
adjoints as in \figfour, with trilinear couplings of the adjoints to
the adjacent bifundamentals replacing (some of) the quartic couplings
of
\quivsup. These trilinear couplings imply that the Ganor strings have
the action $S = \alpha \beta
X$ as above, which upon performing their path integral leads to the
superpotential \polonyi. Thus, the quiver of
\figfour\ can arise from D-branes at the same singularity described in
the previous
subsection.  This raises the possibility of UV completing the SUSY
breaking configuration with a cascading non-Abelian gauge theory.
Then, the Polonyi model would arise as the effective low-energy description
of the SUSY breaking in much the same way that an O'Raifeartaigh model
captures the SUSY breaking vacua of SUSY QCD with slightly massive
quark flavors \IntriligatorDD.
Of course in the spirit of simplicity and minimality, we
are free to consider the final brane configuration of interest (UV
completed by string theory) without invoking the RG cascade and the
consequent increased complexity of our hidden sector.

\ifigure\figfour{The two-node quiver which gives rise directly to a
Polonyi model, after considering the
instanton wrapping symplectic node 1. The arrow from node 2 to itself
is a chiral superfield in the adjoint representation of $U(r_2)$.}  {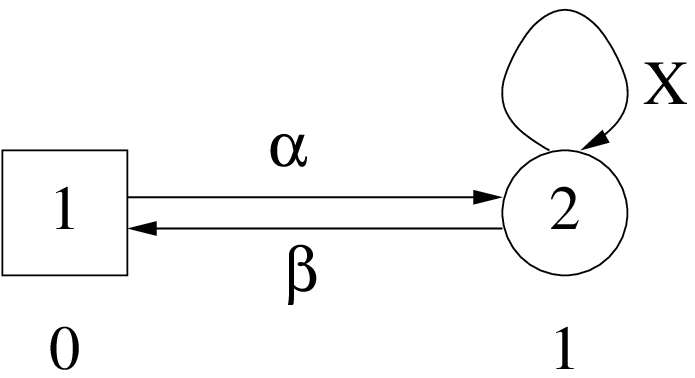}{1.2}

An advantage of obtaining the Polonyi model from
a limit of the Fayet model as above, is that (for suitable $r$) one is
certain of the existence of the stable SUSY breaking minimum; this is
not clear when we obtain the Polonyi model directly from a brane
configuration. It would be interesting to compute the constant $c$ in
the latter case, to see if it leads to a stable SUSY breaking vacuum.

\subsec{An O'Raifeartaigh model}

We obtained the Polonyi model by removing the NS' brane between the
two NS 5-branes in our type IIA brane construction of the Fayet model.
Now, we can make an O'Raifeartaigh model (retrofitted by a stringy
instanton) by inserting another NS-5 brane where the NS' brane
originally was. There are then adjoint fields both for node 2 and for
node 3, as in \tfig\figfive.\foot{Again, one can also obtain this
configuration by performing several steps in the RG cascade of the
theories described in \S2.1 \AharonyPR. Thus,
it corresponds to branes on the same geometrical singularity of \S2.1\ (with
different blow-up parameters).}

\ifigure\figfive{A quiver leading to an O'Raifeartaigh model.}
{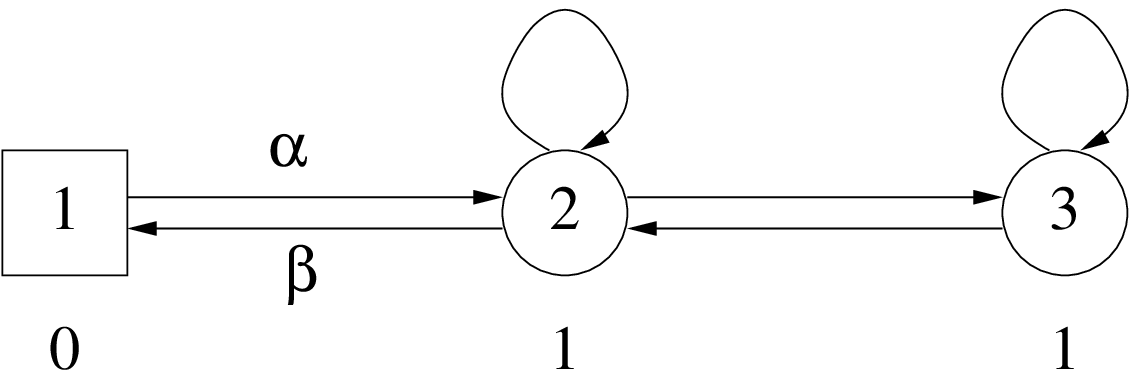}{1.2}

We now have a $U(1) \times U(1)$ gauge group.  Let us call the two
``adjoints'' of $U(1) \times U(1)$ arising at nodes 3 and 2,
respectively, $X$ and $\tilde X$.  In addition, there are
bifundamentals $\Phi, \tilde \Phi$.  The tree-level superpotential is
\eqn\wis{W_{tree} = \tilde \Phi \tilde X \Phi + \tilde \Phi X \Phi~.}
A stringy instanton at node 1 generates a perturbation
\eqn\deltaW{\delta W = \mu^2 \tilde X,} as in \S2.2.  The resulting
full superpotential is
\eqn\wtot{W_{tot} = X \tilde\Phi \Phi + \tilde
X \left( \tilde \Phi \Phi +
\mu^2 \right) ~.}
The $X$ and $\tilde X$ F-terms conspire to break supersymmetry.  In
absence of \deltaW, one could solve the D-term constraint $\vert \phi
\vert^2 - \vert \tilde \phi \vert^2 = r$ by setting one of $\phi,
\tilde \phi$ to $\sqrt{r}$ and the other to zero.
This would yield a supersymmetric vacuum.  The presence of
the stringy instanton effect \deltaW\ instead leads to supersymmetry
breaking, with an exponentially small scale set (in the natural regime
$r \gg \mu^2$) by $\mu$.

This model has a flat direction at this level of analysis.  Lifting
the flat direction by ``UV completing'' the model with a slightly
larger quiver, in analogy with what we did for the Polonyi model in
\S2.2, is one way to potentially stabilize the flat direction.


%
%


\newsec{Discussion}

For realistic model building, there are various options for
communicating supersymmetry breaking to the Standard Model sector.  If
a Standard Model brane system sits far away from our SUSY-breaking
system, we may obtain gravity mediation.  We can also generalize the
models above in a straightforward way to obtain messengers appropriate
for gauge mediation, as
follows.\foot{For a review of
gauge mediation, see \GiudiceBP.  For recent attempts
to engineer such models using branes, see \refs{\DiaconescuPC,
\GarciaEtxebarriaRW,\KawanoRU,
\AmaritiAM}.}
Consider (for example) the extension
of the brane system of \S2.1\ depicted in
\tfig\figsix, where we now occupy node 4 with a toy ``Standard Model."
This introduces a second set of chiral fields $\eta, \tilde \eta$
charged under the new gauge group, and a superpotential of the form
\eqn\Wwithm{W =\Lambda_1 \Phi_+ \Phi_- +
{1\over M_*} \eta \tilde \eta \Phi_+ \Phi_- + M \eta\tilde\eta, }
where the quartic term arises from the superpotential \quivsup, and
we have included a possible supersymmetric mass term $M$ for
$\eta,\tilde\eta$.
In the supersymmetry breaking vacuum with $\phi_+\sim\sqrt{r}$ and
$F_{\Phi_-}\sim \Lambda_1 \sqrt{r}$, the operator $\Phi_+\Phi_-$ has
zero VEV and an $F$ component of order $\langle \phi_+ \rangle
F_{\Phi_-}$. As a result, the superpotential \Wwithm\ is of the form
appropriate for gauge mediation with messengers $\eta,\tilde\eta$ of
mass $M$, and with an effective SUSY-breaking F-term of order $\langle
\phi_+ \rangle F_{\Phi_-}/M_*\sim r \Lambda_1/M_*$.  The quartic term
in \Wwithm\ leads to the existence of additional (supersymmetric) vacua
far away in field space, but it does not affect the non-supersymmetric
vacuum that we are interested in (which is now metastable).


\ifigure\figsix{A quiver with a coupling to the ``Standard Model'' at
node 4 and symplectic nodes (with stringy D-instantons) at nodes 1 and 5.}
{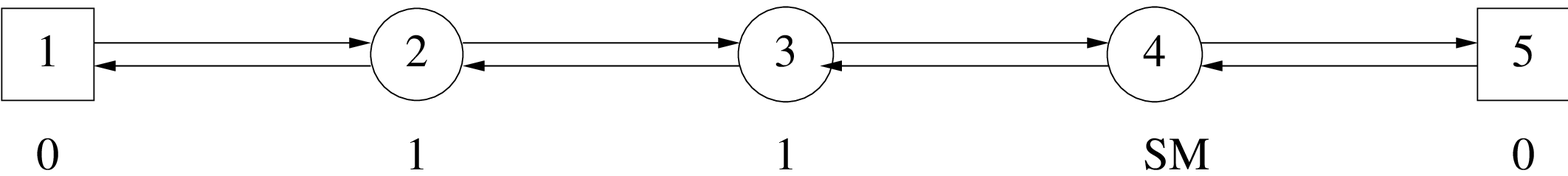}{0.8}

For high-scale gauge mediation, one requires a messenger mass $M$ well
below the string scale but much higher than the TeV scale.\foot{This
has various phenomenological advantages \refs{\IbeKM,\Feng}.} One possibility for
obtaining such a mass is by turning on closed string moduli (blow-up
modes), and this then involves a small tune of parameters.  If one
prefers a dynamical mechanism to obtain $M$, which is particularly
important for lower-scale gauge mediation, one can (as in \figsix)
make node 5 another (unoccupied) symplectic node. Then, if we put a
single brane at node 4, we get a mass term for $\eta, \tilde
\eta$ of magnitude $\Lambda_5$ from the stringy instanton at node 5.
This provides a tunable messenger mass.
Since node 4 must be a $U(1)$ for this to happen, we would need to
consider an extension of the Standard Model by this $U(1)$ symmetry,
with appropriate charges to get gauge-mediated masses from this setup.

In order to obtain a realistic model, we could investigate the
possibility of replacing node 4 above with a full brane realization of
the Standard Model (rather than the toy version described above), which is
(classically) mutually supersymmetric with our SUSY-breaking sector.
In doing so we must require that the new open strings $\eta,
\tilde \eta$ connecting our SUSY-breaking theory to the Standard Model
have the couplings \Wwithm\ (which are the
lowest order couplings allowed by the gauge symmetries). Again, one
would need to generate messenger masses by an appropriate choice of
closed string moduli, or by a dynamical mechanism similar to the one
described above. It would be interesting to construct an explicit
model of this sort, and to explore to what extent our simple DSB
sectors (or obvious analogues) can easily be incorporated in existing
semi-realistic brane constructions of the Standard Model (such as
\refs{\BlumenhagenMU,\VerlindeJR}).
It would also be worthwhile to find analogous DSB models in the limits of
string theory which more readily admit unification of coupling
constants.  Another natural generalization may be to apply
D-instanton retrofitting to the recently studied O'Raifeartaigh models
which spontaneously break R-symmetry \ShihAV.

\medskip
\centerline{\bf{Acknowledgements}}

This work is supported in part by the Israel-U.S. Binational Science
Foundation.  OA is supported by a center of excellence supported by
the Israel Science Foundation (grant number 1468/06), by the European
network HPRN-CT-2000-00122, by a grant from the G.I.F., the
German-Israeli Foundation for Scientific Research and Development, and
by a grant of DIP (H.52). OA thanks the Isaac Newton Institute for
Mathematical Sciences for hospitality during the completion of this
work. The research of SK and ES is supported in part by NSF grant
PHY-0244728, and in part by the DOE under contract DE-AC03-76SF00515.

\listrefs
\end